\documentclass[floatfix,pra,twocolumn,showpacs,amsmath,amssymb,letterpaper,groupaddresses,superscriptaddress]{revtex4}
\usepackage{times}
\usepackage{latexsym}
\usepackage{graphicx}
\usepackage{verbatim,times,bbm}
\usepackage{color}

\usepackage{color}

\def\be{\begin{equation}}
\def\ee{\end{equation}}
\def\ba{\begin{eqnarray}}
\def\ea{\end{eqnarray}}

\def\ra{\rangle}

\def\A1{A_{-1}}

\begin{document}

\title{Stationary entanglement achievable by environment induced chain links}

\author{Laleh Memarzadeh\footnote{email: memarzadeh@sharif.edu}}
\affiliation{School of Science and Technology, University of Camerino, I-62032 Camerino, Italy}
\affiliation{Department of Physics, Sharif University of Technology, Teheran, Iran}

\author{Stefano Mancini\footnote{email: stefano.mancini@unicam.it}}
\affiliation{School of Science and Technology, University of Camerino, I-62032 Camerino, Italy}
\affiliation{INFN-Sezione di Perugia, I-06123 Perugia, Italy}

\begin{abstract}
We investigate the possibility of chaining qubits by letting pairs of nearest neighbours qubits dissipating into common environments. We then study entanglement dynamics within the chain and show that steady state entanglement can be achieved.
\end{abstract}


\maketitle

\section{Introduction}

It is nowadays well established that entanglement represents a fundamental resource for quantum information tasks \cite{vedral07}. However, being a purely quantum feature it is fragile with respect to enviromental contamination.
Notwithstanding that, the possibility to achieve entangled states as stationary ones  in open quantum systems has been put forward in many different contexts (for what concern qubits systems see e.g. Refs.\cite{braun02,clark03}). The subject has attracted a lot of attention up until a recent striking experiment on long living entanglement \cite{polzik10}.
The works on this topic can be considered as falling into two main categories:
one where all qubits are plunged in the same environment \cite{braun02} and the other where each qubit is plunged in its own environment \cite{clark03}.
In the former case the environment can provide an indirect interaction between otherwise decoupled qubits and thus a means to entangle them.
In the latter case a direct interaction between qubits is needed to create entanglement, and usually to maintain it one has to also exploit other mechanisms (state resetting, driving, etc.).

Here we consider a hybrid situation as depicted in Fig.1.
It represents a sort of spin-$\frac{1}{2}$ chain dimerized by environments.
In practice each environment induces a chain link between contiguous qubits.
Hence, we can expect that a simple dissipative dynamics in such a configuration is able to establish entanglement along the chain without the need to exploit any other mechanism.
Actually, we will show for the case of three qubits the possibility of achieving stationary entanglement for each qubits pair. The amount of entanglement results strongly dependent on the initial (separable) state. Also the dependance from the chain boundary conditions (open or closed) will be analyzed as well as a left-right asymmetry in qubit-environment interaction.

The layout of the paper is the following:
in Section II we introduce the model relying on physical motivations and we discuss the general dynamical properties;
in Section III we restrict our attention to the three qubits case and investigate the entanglement dynamics in the open boundary condition;
in Section IV we analyze the same system but with closed boundary conditions.
Concluding remarks are presented in Section V.

%%%%%%%%%%%%%%%%%%%%%%%%%%%%%%%%%%%

\section{The Model}

\begin{figure}[t]
\centering
\vspace{-0cm}
\includegraphics[scale=0.35]{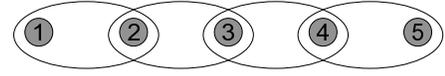}
\vspace{-2cm}
\caption{Environments (ellipses) inducing chain links between contiguous qubits (gray circles).}
\label{chain}
\end{figure}

The model of Fig.1 can be motivated by physically considering two-level atoms inside cavities connected by fibers \cite{serafini06}. In such a scheme each atom-qubit can be thought as exchanging energy with the optical modes supported by the fiber. In turn this latter can modeled as an environment.
Thus each qubit dissipates energy through two environments (one on the left and the other on the right).
It happens that two contiguous qubits dissipates energy into the same environment. Then this environment mediates the interaction between the contiguous qubits.

More specifically, let us consider at the $i$th site of a chain a qubit described by ladder operators $\hat{\sigma}^{\pm}_i$ satisfying the usual spin-$\frac{1}{2}$ algebra  $[\hat{\sigma}^{+}_i,\hat{\sigma}^{-}_i]=\hat{\sigma}_i^z$.
Let us also consider at the $i$th site of a chain radiation modes described by ladder operators $\hat{b}_{i,j},\hat{b}^{\dag}_{i,j}$ satisfying the usual bosonic algebra $[\hat{b}_{i,j},\hat{b}^{\dag}_{i,j^\prime}]=\delta_{j,j^\prime}$.
Then, the interaction Hamiltonian reads
\begin{equation}
\hat{H}_I=\sum_i\sum_j \hat{\sigma}_{i}^{-}\left(\hat{b}_{i-1,j}^{\dag}+\hat{b}_{i,j}\right)+\textrm{h.c.}.
\end{equation}
By considering the $\hat{b}_{i,j}$ as environment's operators for the  $i$th qubit, we can use
standard techniques \cite{qnoise} to arrive at the following master equation
\begin{eqnarray}\label{Dyn}
\frac{\partial\hat{\rho}}{\partial t}&=&\sum_{i}\left[(\hat{\sigma}_i^-+\hat{\sigma}_{i+1}^-)\hat{\rho} (\hat{\sigma}_i^++\hat{\sigma}_{i+1}^+)\right.\nonumber\\
&&\left.-\frac{1}{2}\left\{(\hat{\sigma}_i^++\hat{\sigma}_{i+1}^+)(\hat{\sigma}_i^-+\hat{\sigma}_{i+1}^-),\hat{\rho}\right\}\right],
\end{eqnarray}
where $\{,\}$ denotes the anti-commutator and we have assumed unit decay rate.

Since we are interested on the steady state we have to notice that,
given a master equation written in the standard Linbladian form,
\begin{equation}
\frac{\partial\hat{\rho}}{\partial t}=\sum_{i}\left[\hat{L}_i\hat{\rho} \hat{L}_i^{\dag}-\frac{1}{2}\left\{\hat{L}_i^{\dag}\hat{L}_i,\hat{\rho}\right\}\right],
\end{equation}
the uniqueness of the stationary solution is guaranteed if the only operators commuting with every Lindblad operator $\hat{L}_i$ are multiples of identity \cite{Spohn}.

In the case of Eq.\eqref{Dyn} the $\hat{\sigma}_i^{-}$s commute with Lindblad operators. Hence the steady state may not be unique, that is it may depend on the initial conditions. Due to that we need to study the full dynamics of the system.

%%%%%%%%%%%%%%%%%%%%%%%%%%%%%%%%%%%

\section{The three qubits case with open boundary conditions}

We restrict our attention to a chain of three sites. We first consider open boundary conditions.
Then, the dynamics will be described by a master equation that can be easily derived from Eq.(\ref{Dyn})
\begin{eqnarray}\label{Dyn1}
\frac{\partial\hat{\rho}}{\partial t}&=&\gamma\left[2(\hat{\sigma}_1^-+\hat{\sigma}_{2}^-)\hat{\rho} (\hat{\sigma}_{1}^{+}+\hat{\sigma}_{2}^{+})\right.\nonumber\\
&&\left.-\{(\hat{\sigma}_{1}^{+}+\hat{\sigma}_{2}^{+})(\hat{\sigma}_{1}^-+\hat{\sigma}_{2}^-),\hat{\rho}\}\right]\nonumber\\
&+&(1-\gamma)\left[2(\hat{\sigma}_2^-+\hat{\sigma}_{3}^-)\hat{\rho} (\hat{\sigma}_{2}^{+}+\hat{\sigma}_{3}^{+})\right.\nonumber\\
&&\left.-\{(\hat{\sigma}_{2}^{+}+\hat{\sigma}_{3}^{+})(\hat{\sigma}_{2}^-+\hat{\sigma}_{3}^-),\hat{\rho}\}\right].
\end{eqnarray}
Here we have considered the possibility for each qubit of having an asymmetric decay rate on the left and right environments. This has been accounted for by the real factors $\gamma$ and $(1-\gamma)$ with the assumption $0<\gamma<1$.
Clearly the symmetric situation is recovered when $\gamma=1/2$.

By arranging the density matrix (expressed in the computational basis $\{|e\rangle,|g\rangle\}^{\otimes 3}$)
as a vector $\mathbf{v}$ (e.g. writing $\rho_{i,j}={\mathbf v}_{8(i-1)+j}$),
the master equation \eqref{Dyn1} can be rewritten as a linear set of differential equations
\begin{equation}
\dot{\mathbf{v}}(t)=M{\mathbf{v}}(t),
\label{lset}
\end{equation}
where $M$ is a $64\times 64$ matrix of constant coefficients given by
\begin{eqnarray}
M&=&\gamma(2L_1\otimes L_1-L_1^{\dag}L_1\otimes I_8-I_8\otimes L_1^{\dag}L_1)\nonumber\\
&+&(1-\gamma)(2L_2\otimes L_2-L_2^{\dag}L_2\otimes I_8-I_8\otimes L_2^{\dag}L_2),
\end{eqnarray}
where
\begin{eqnarray}
L_1&=&\sigma^-\otimes I_2\otimes I_2+I_2\otimes \sigma^-\otimes I_2,\nonumber\\
L_2&=&I_2\otimes\sigma^-\otimes I_2+I_2\otimes I_2\otimes \sigma^-,
\end{eqnarray}
with $I_n$ the $n\times n$ dimensional identity matrix and $\sigma^+=\left(\begin{array}{cc}0&1\\0&0\end{array}\right)$, $\sigma^-=\left(\begin{array}{cc}0&0\\1&0\end{array}\right)$.
Then, the set of differential equations \eqref{lset} can be converted into a set of algebraic equations via the Laplace transform,
$\tilde{\mathbf{v}}(s)=\int_0^{\infty}\exp(-st){\mathbf{v}}(t)$, i.e.
\begin{equation}
s\tilde{\mathbf{v}}(s)-{\mathbf{v}}(0)=M\tilde{\mathbf{v}}(s).
\end{equation}
Decoupling these equations one finds that the Laplace transforms of the density matrix elements are rational functions of
polynomials and the inverse Laplace transformation can be performed analytically.
The results are not explicitly reported because the expressions are too much cumbersome.

Having the density matrix of the system, we can study the entanglement dynamics for each qubit pair of the system.
To quantify the amount of entanglement between each of the qubits we use the concurrence \cite{Wootters}. We recall that to find the concurrence of a bipartite system described by the density matrix $\rho$, the following steps
should be done:
\begin{itemize}
\item[1)] Find the complex conjugate of the density matrix in the computational basis and denote it by  $\rho^*$.
\item[2)] Define $\tilde{\rho}:=(\sigma_y\otimes\sigma_y)\rho^*(\sigma_y\otimes\sigma_y)$, where $\sigma_y=i(\sigma^--\sigma^+)$.
\item[3)] Find the square root of the eigenvalues of $\rho\tilde{\rho}$ and sort them in decreasing order: $\{\lambda_1,\lambda_2,\lambda_3,\lambda_4\}$.
\item[ 4)] The concurrence is given by
\begin{equation}
C=\max \{0,\lambda_1-\lambda_2-\lambda_3-\lambda_4\}.
\end{equation}
\end{itemize}

\subsection{Entanglement dynamics}

\begin{figure}[t]
\centering
\includegraphics[scale=0.45]{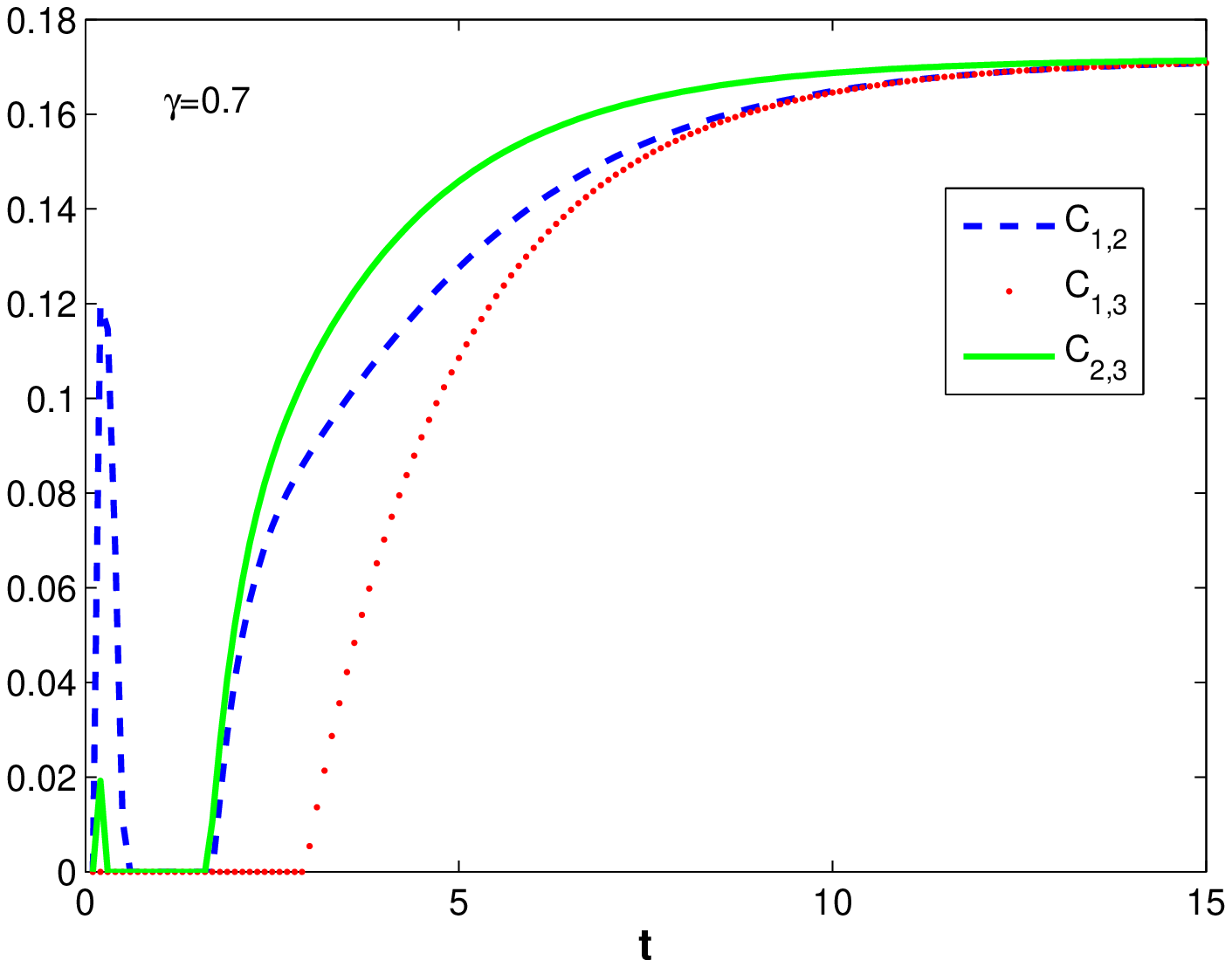}
\caption{Qubit pair concurrence vs time for $|eee\ra$ initial state. Dashed line: $C_{_{1,2}}$, solid line: $C_{_{2,3}}$ and
dotted line: $C_{_{1,3}}$.}\label{3e}
\end{figure}

Figure \ref{3e} shows the evolution of entanglement between each qubit pair for $|eee\ra$ initial state.
As it can be seen in this figure, when all the qubits are initially in excited state it takes longer time for the first and the third
qubits to become entangled compared to the time needed to generate entanglement between the first and second or second and third qubits. As a consequence for not nearest neighborhood qubits we have a \emph{sudden birth of entanglement}, i.e. it suddenly becomes non zero at times greater than zero (it does not smoothly increases starting from initial time) \cite{FT08}.

%%%%%%%%%%%%%%%%%%%%%%%%%%%%%%%%%%%%%%%%
%%%%%%%%%%%%%%%%%%%%%%%%%%%%%%%%%%%%%%%%%%%%%%%%%

\begin{figure}[t]
\centering
\includegraphics[scale=0.45]{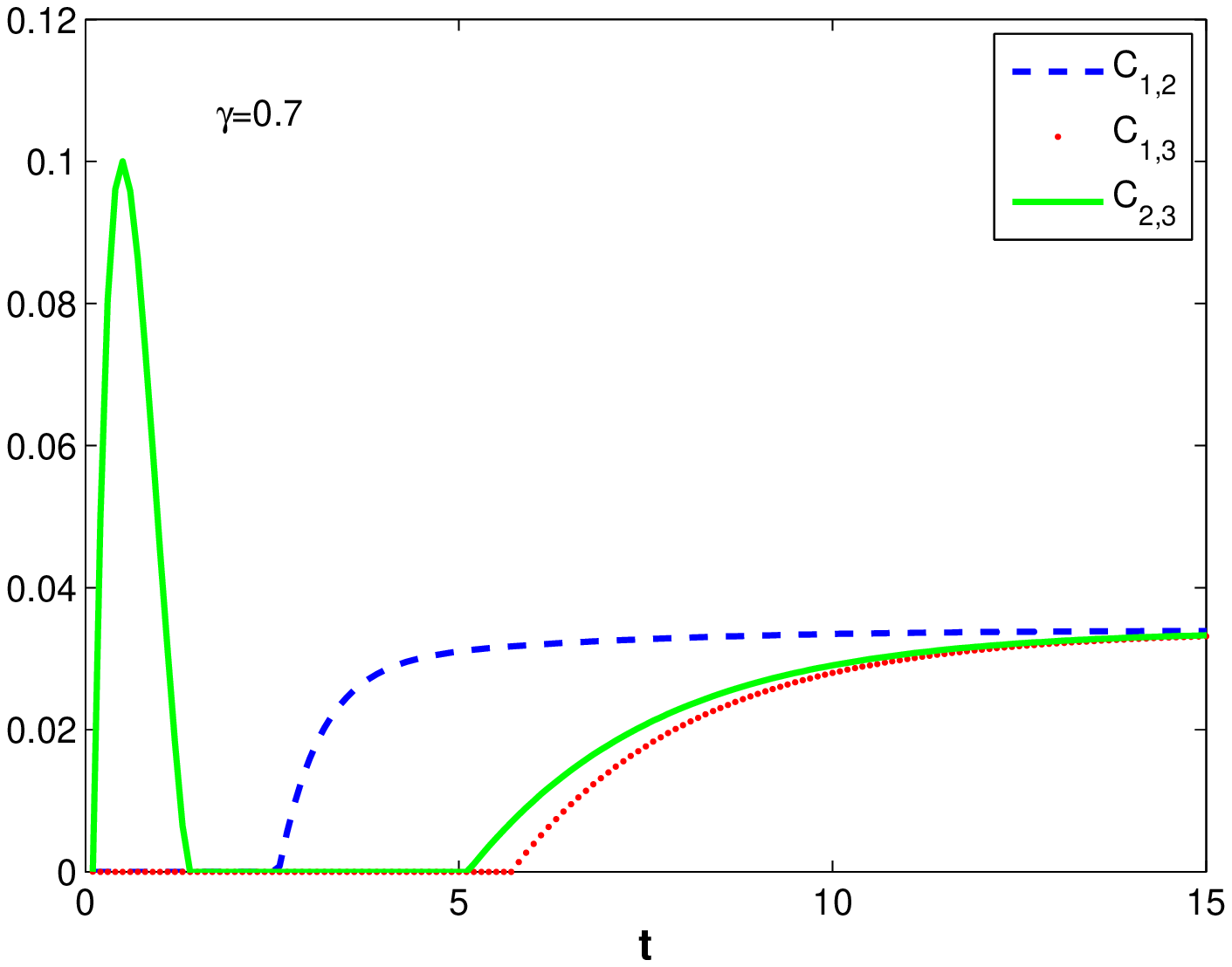}\quad\includegraphics[scale=0.45]{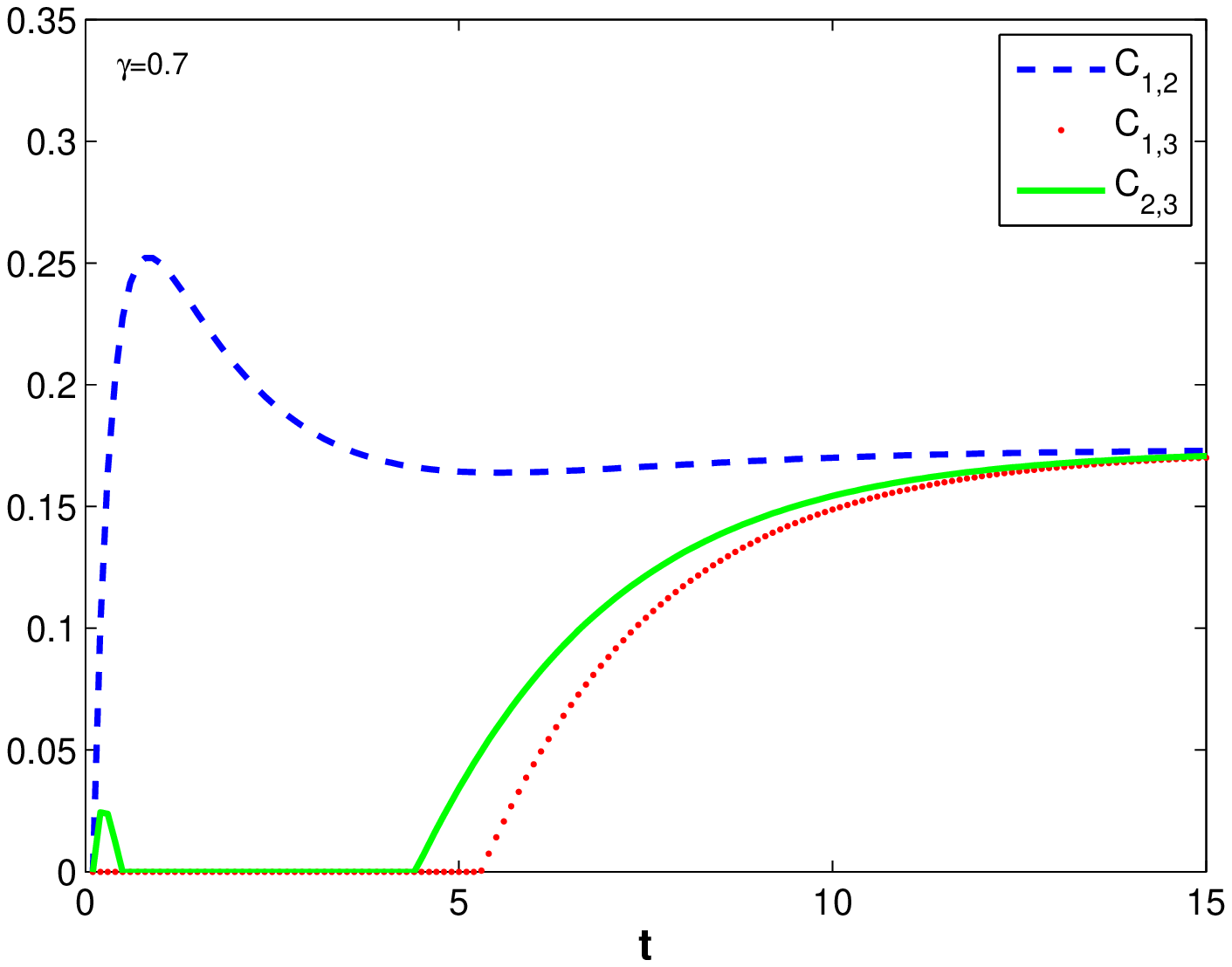}
\caption{Qubit pair concurrence vs time for $|eeg\ra$ (top) and $|ege\rangle$ (bottom) initial states. Dashed line: $C_{_{1,2}}$, solid line: $C_{_{2,3}}$ and
dotted line: $C_{_{1,3}}$.}
\label{2e}
\end{figure}

If we start with different initial states, the entanglement behaves differently in time. Figure \ref{2e}-left shows the value of concurrence in time for $|eeg\ra$ as initial state.
Like the previous case, it shows that the time needed by the first and third qubits to become entangled is longer than the time needed by nearest neighborhood qubits. Moreover, we still have the entanglement sudden birth phenomenon, but this time
it manifests not only for distant qubits (first and third) but also for the first and the second qubits.
The interesting point in this figure is that the entanglement generation between the nearest neighbors is quicker
if they are initially prepared in $|eg\ra$ rather than $|ee\ra$ state.
The other possibility with two number of excitations in the initial state is $|ege\ra$, for which the time evolution of entanglement in shown in Figure \ref{2e}-right. This time the entanglement sudden birth phenomenon only manifests for distant qubits (first and third).

%%%%%%%%%%%%%%%%%%%%%%%%%%%%%%%%%%%%%%%%%%%%%%%%%
%%%%%%%%%%%%%%%%%%%%%%%%%%%%%%%%%%%%%%%%%%%%%%%%%

%%%%%%%%%%%%%%%%%%%%%%%%%%%%%%%%%%%%%%%%%%%%%%%%%
\begin{figure}[t]
\centering
\includegraphics[scale=0.45]{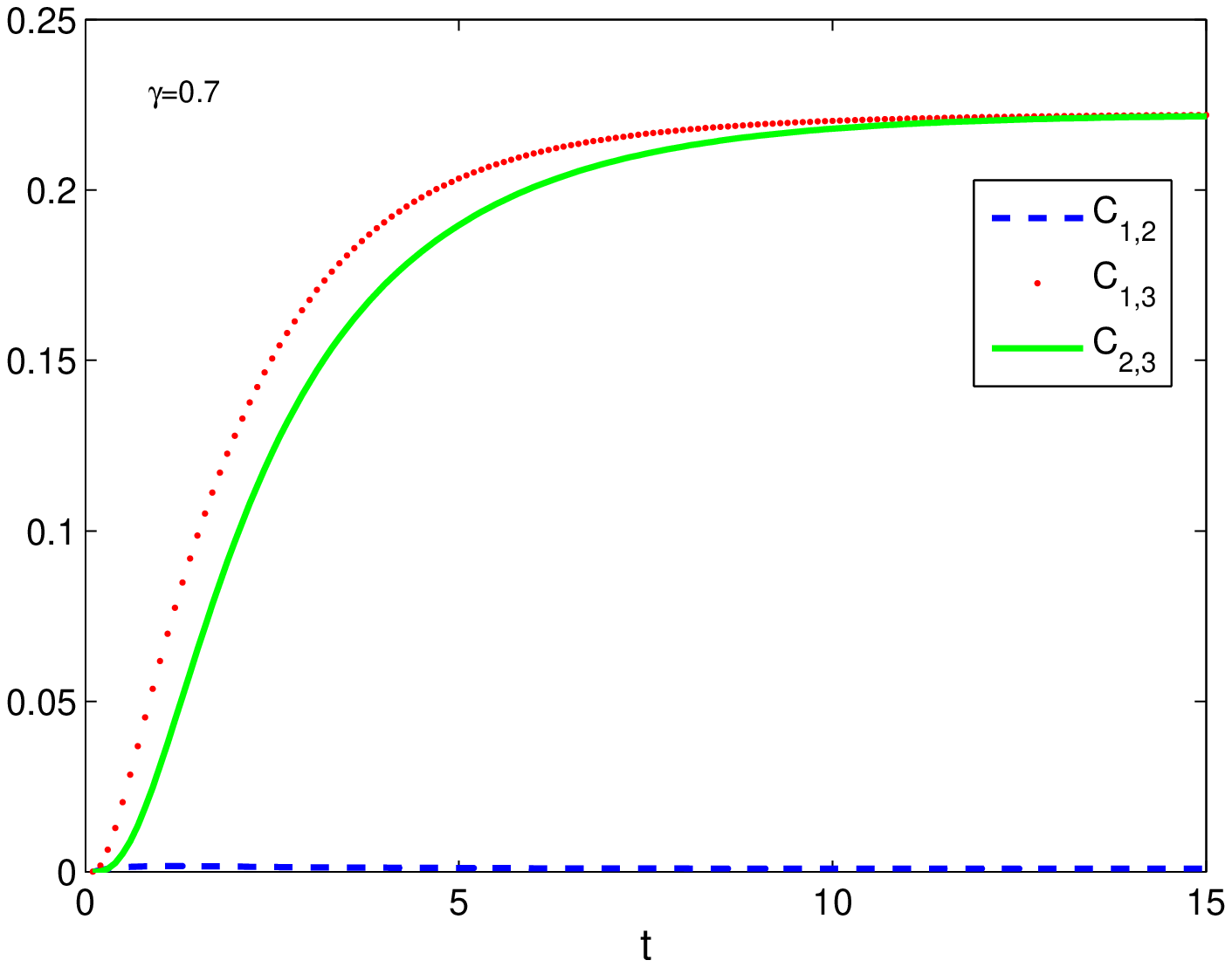}\quad\includegraphics[scale=0.45]{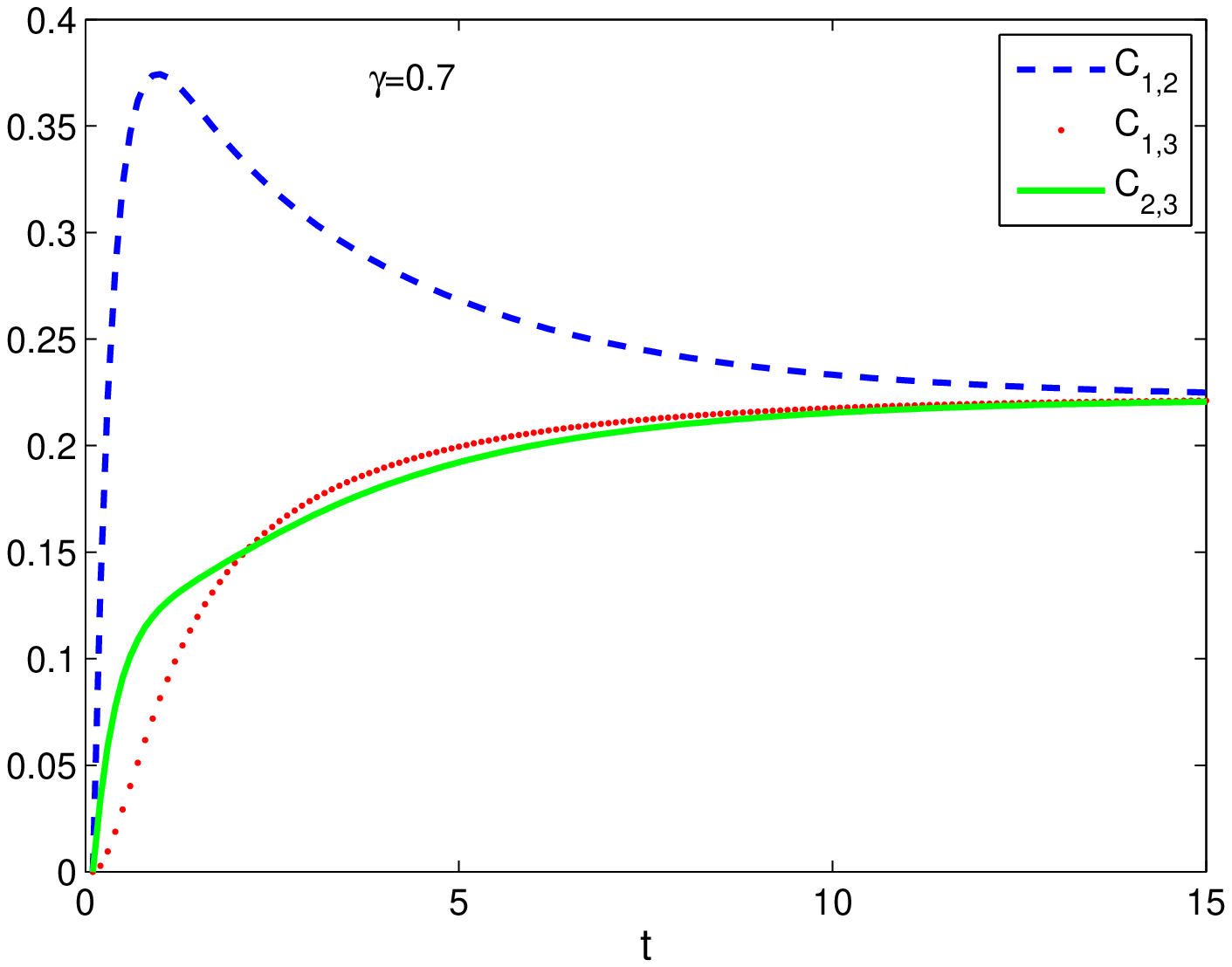}
\caption{Qubit pair concurrence vs time for $|egg\ra$ (top) and $|geg\rangle$ (bottom) initial states. Dashed line: $C_{_{1,2}}$, solid line: $C_{_{2,3}}$ and
dotted line $C_{_{1,3}}$.}
\label{1e}
\end{figure}
%%%%%%%%%%%%%%%%%%%%%%%%%%%%%%%%%%%%%%%%%%%%%%%%%

A big difference appears if the number of excitation of the initial state, reduces to one. Figures \ref{1e}-left and-right show entanglement evolution
for the initial states $|egg\ra$ and $|geg\ra$, respectively. As it can be seen in this figure, entanglement is generated between each qubits pair from the beginning, no matter how far they are from each other, i.e. we no more have entanglement sudden birth phenomenon.

Finally, in the case of $|ggg\ra$ initial state there is no entanglement at any time because this state represent a fixed point of the Liouvillian superoperator at right hand side of the master equation \eqref{Dyn1}, or in other words $M\mathbf{v}=0$ with $\mathbf{v}_j=\delta_{j,64}$.

\subsection{Stationary entanglement}

Taking the limit $t\to \infty$ in the density matrix elements, we arrive at
the following general form for the steady state
\begin{equation}
\rho_s=\left(\begin{array}{cccccccc}
0&0&0&0&0&0&0&0\cr
0&0&0&0&0&0&0&0\cr
0&0&0&0&0&0&0&0\cr
0&0&0&f&0&-f&f&0\cr
0&0&0&0&0&0&0&0\cr
0&0&0&-f&0&f&-f&0\cr
0&0&0&f&0&-f&f&0\cr
0&0&0&0&0&0&0&1-3f\cr
\end{array}\right),
\end{equation}
where $f\in\mathbb{R}$ is determined by the initial state of the system. In particular we have the following correspondence:
\begin{eqnarray}
f &=& \frac{24-19\gamma+19\gamma^2}{216 + 81\gamma-81\gamma^2},  \quad |eee\rangle \nonumber\\
f &=& \frac{4(4- 5\gamma +\gamma^2)}{
27(8 + 3\gamma -3\gamma^2)},  \hskip 5mm |eeg\rangle \nonumber\\
f &=& \frac{4(4+ 5\gamma -5\gamma^2)}{
27(8 + 3\gamma -3\gamma^2)},  \hskip 5mm |ege\rangle \nonumber\\
f &=& \frac{1}{9}, \hskip 21mm |egg\rangle,|geg\rangle,|gge\rangle   \nonumber\\
f &=& \frac{4\gamma(3+ \gamma)}{
27(8 + 3\gamma -3\gamma^2)},  \hskip 5mm |gee\rangle \nonumber\\
f &=& 0,  \hskip 32mm |ggg\rangle.
\end{eqnarray}
To find the amount of entanglement in each qubits pair at the steady state, we first write the reduced density matrices
\begin{equation}\label{reduced}
\rho_{_{1,2}}=\rho_{_{2,3}}=\left(\begin{array}{cccc}
0&0&0&0\cr
0&f&-f&0\cr
0&-f&f&0\cr
0&0&0&1-2f\end{array}\right),
\end{equation}
\begin{equation}
\rho_{_{1,3}}=\left(\begin{array}{cccc}
0&0&0&0\cr
0&f&f&0\cr
0&f&f&0\cr
0&0&0&1-2f\end{array}\right).
\end{equation}
%%%%%%%%%%%%%%%%%%%%%%%%%%%%%%%%%%%%%%%%
\begin{figure}[t]
\centering
\includegraphics[scale=0.45]{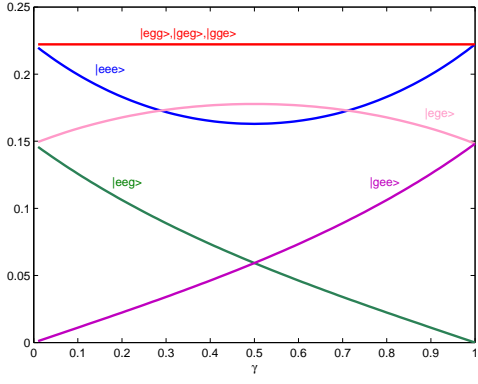}
\caption{Steady state entanglement versus $\gamma$ for different initial states.}\label{MathematicaSteady}
\end{figure}
%%%%%%%%%%%%%%%%%%%%%%%%%%%%%%%%%%%%%%%%%%%%%%%%%
Then, it is easy to show that the concurrence becomes
\begin{equation}\label{Copen}
C_{12}=C_{13}=C_{23}=2f.
\end{equation}
Figure (\ref{MathematicaSteady}) shows the stationary entanglement vs $\gamma$ for different initial state.
For initial states like $|egg\rangle, |geg\rangle,  |gge\rangle$ the entanglement does not depend on $\gamma$ .
For $|eee\rangle$ initial state, by increasing the difference in the dissipation
rates more entanglement will be induced in the system. On the contrary for $|ege\rangle$
the maximum value of entanglement is achieved when the dissipation rates into the two environments are
the same. As it is expected the entanglement for initial states
$|eeg\rangle, |gee\rangle$ are equal at $\gamma = 1/2$.
In these cases, the maximum value can be attained when
there is one excitation in the initial state.

%%%%%%%%%%%%%%%%%%%%%%%%%%%%%%%%%%%%%%%

\section{The three qubits case with closed boundary conditions}

We now study the chain of three sites with closed boundary condition. In this case the master equation \eqref{Dyn} reads
\begin{eqnarray}
\frac{\partial\hat{\rho}}{\partial t}&=&\gamma\left[2(\hat{\sigma}_1^-+\hat{\sigma}_{2}^-)\hat{\rho} (\hat{\sigma}_{1}^{+}+\hat{\sigma}_{2}^{+})\right.\nonumber\\
&&\left.-\{(\hat{\sigma}_{1}^{+}+\hat{\sigma}_{2}^{+})(\hat{\sigma}_{1}^-+\hat{\sigma}_{2}^-),\hat{\rho}\}\right]\nonumber\\
&+&\mu\left[2(\hat{\sigma}_2^-+\hat{\sigma}_{3}^-)\hat{\rho} (\hat{\sigma}_{2}^{+}+\hat{\sigma}_{3}^{+})\right.\nonumber\\
&&\left.-\{(\hat{\sigma}_{2}^{+}+\hat{\sigma}_{3}^{+})(\hat{\sigma}_{2}^-+\hat{\sigma}_{3}^-),\hat{\rho}\}\right]\nonumber\\
&+&\nu\left[2(\hat{\sigma}_3^-+\hat{\sigma}_{1}^-)\hat{\rho} (\hat{\sigma}_{3}^{+}+\hat{\sigma}_{1}^{+})\right.\nonumber\\
&&\left.-\{(\hat{\sigma}_{3}^{+}+\hat{\sigma}_{1}^{+})(\hat{\sigma}_{3}^-+\hat{\sigma}_{1}^-),\hat{\rho}\}\right],
\label{mecc}
\end{eqnarray}
where the possibility for each qubit of having an asymmetric decay rate on the left and right environments is accounted
for by the real factors $\gamma$, $\mu$ and $\nu$ ($0\le\gamma, \mu, \nu\le 1$).
Clearly the symmetric situation is recovered when $\gamma=\mu=\nu$ ($=1/2$).

Again arranging the density matrix (expressed in the computational basis $\{|e\rangle,|g\rangle\}^{\otimes 3}$)
as a vector $\mathbf{v}$ (e.g. writing $\rho_{i,j}={\mathbf v}_{8(i-1)+j}$),
the master equation \eqref{mecc} can be rewritten as a linear set of differential equations
\begin{equation}
\dot{\mathbf{v}}(t)=M{\mathbf{v}}(t),
\end{equation}
where $M$ is a $64\times 64$ matrix of constant coefficients given by
\begin{eqnarray}
M&=&\gamma(2L_1\otimes L_1-L_1^{\dag}L_1\otimes I_8-I_8\otimes L_1^{\dag}L_1)\nonumber\\
&+&\mu(2L_2\otimes L_2-L_2^{\dag}L_2\otimes I_8-I_8\otimes L_2^{\dag}L_2)\nonumber\\
&+&\nu(2L_3\otimes L_3-L_3^{\dag}L_3\otimes I_8-I_8\otimes L_3^{\dag}L_3),
\end{eqnarray}
where
\begin{eqnarray}
L_1&=&\sigma^-\otimes I_2\otimes I_2+I_2\otimes \sigma^-\otimes I_2,\nonumber\\
L_2&=&I_2\otimes\sigma^-\otimes I_2+I_2\otimes I_2\otimes \sigma^-,\nonumber\\
L_2&=&I_2\otimes I_2 \otimes\sigma^-+\sigma^-\otimes I_2\otimes I_2.
\end{eqnarray}
In this case it is possible to see that
\begin{equation}
M{\mathbf{v}}=0,
\end{equation}
admits a unique solution ${\mathbf{v}}_j=\delta_{j,64}$, i.e. the only possible solution is $|ggg\rangle$,
 for all values of $\gamma,\mu,\nu$. Hence no entanglement survive at stationary conditions.

%%%%%%%%%%%%%%%%%%%%%%%%%%%%%%%%%%%%%%%

\section{Conclusion}

In this work we have considered a spin-$\frac{1}{2}$ chain dimerized by environments.
By means of dissipative mechanism, each environment induces a chain link between contiguous qubits.
Then we have studied the possibility of having long living entanglement without resorting to any other mechanism.
In particular for the case of three qubits chain with open boundary condition we have classified the amount of stationary entanglement accordingly to some initial (separable) states.
Here we have also shown the appearance of  the entanglement sudden birth.
On the contrary, for the case of three qubits chain with closed boundary condition we have proved the impossibility of stationary entanglement.
This fact can be interpreted as entanglement frustration phenomenon \cite{Facchi}, induced in this context by the imposed periodic boundary conditions.

The proposed scheme lends itself to to be extended to $n>3$ sites where one can evaluate
how entanglement scales as function of the distance between the two qubits.
Moreover, in the limit of large $n$, it could also results useful for studying possible connections
between quantum phase transitions and reservoirs properties.

Finally, the discussed method of generating entanglement seems economical and offers interesting perspectives
for the generation of the so called graph states (useful for measurement based quantum compuatation) \cite{Rauss},
when one considers a network topologies more complicate than a simple chain.

\acknowledgments 
We acknowledge the financial support of the European Commission, under the FET-Open grant agreement HIP, number FP7-ICT-221889.

\end{document}